\documentclass[9pt]{article}

\usepackage{listings}
\usepackage{lineno}

\usepackage[utf8]{inputenc}
\usepackage[english]{babel}
\usepackage{amssymb}
\usepackage{amsmath}
\usepackage{graphicx}
\graphicspath{ {./images/} }
\usepackage[export]{adjustbox}
\usepackage{subcaption}

\usepackage{pgfplotstable}
\usepackage{siunitx}

\DeclareMathOperator*{\argmax}{arg\,max}

\usepackage[margin=0.75in, top=0.75in]{geometry}

\usepackage{framed}
\usepackage{authblk}

\usepackage{enumitem}
\newlist{todolist}{itemize}{2}
\setlist[todolist]{label=$\square$}

\usepackage[
backend=biber,
style=alphabetic,
citestyle=authoryear
]{biblatex}
\usepackage[colorlinks,citecolor=blue,urlcolor=blue,bookmarks=false,hypertexnames=true]{hyperref}

\title{Computational Causal Inference}

\author{
  Jeffrey C. Wong\\
  Computational Causal Inference, Netflix
}

\date{}

\begin{document}

\maketitle

\begin{abstract}
    We introduce computational causal inference as an interdisciplinary field across causal inference, algorithms design and numerical computing. The field aims to develop software specializing in causal inference that can analyze massive datasets with a variety of causal effects, in a performant, general, and robust way. The focus on software improves research agility, and enables causal inference to be easily integrated into large engineering systems. In particular, we use computational causal inference to deepen the relationship between causal inference, online experimentation, and algorithmic decision making.
    
    This paper describes the new field, the demand, opportunities for scalability, open challenges, and begins the discussion for how the community can unite to solve challenges for scaling causal inference and decision making.
\end{abstract}

\section{Introduction}

Causal inference and machine learning have a symbiotic relationship that is growing deeper. Companies are using machine learning to improve content recommendations, sales, business operations, and to personalize user experiences. These companies will test new algorithms online in order to determine whether the algorithms cause a positive effect for the company. In this capacity, causal inference for online experiments serves as an honest and independent evaluator for an algorithm.
However, recent interdisciplinary work in the combination of machine learning and causal inference has shown a much deeper synergy between the two fields.
Predictions from machine learned models have been debiased by utilizing inverse propensity weights (\cite{dudik2011doubly}), which are frequently found in studies of causal effects. At the same time, causal inference methods have benefited from methods for modeling high dimensional relationships in order to determine heterogeneity in treatment effects, such as in \cite{wager2018estimation}. Frameworks such as Pearl's do-calculus (\cite{pearl2012calculus}) have also created clear programmatic structure for answering causal effects queries when relationships in data can be modeled as a graph.

The intersection of machine learning and causal inference is particularly strong in the symbiotic relationship between algorithmic policy making and experimentation platforms. In policy making, we are presented with a decision to make; we wish to construct an algorithm that receives features as input and outputs an action to take. The action can be personalized, for example in contextual bandits (\cite{li2010contextual}), and should be the optimal action that maximizes a reward function. These algorithms are tested in an online experiment that reports the causal effect on a key performance indicator (KPI) due to the new algorithm. Experimentation and policy making are highly aligned when the reward function for the policy algorithm and the KPI used for an online experiment are the same, thus the policy algorithm must determine the action with the largest causal effect on the KPI. Recent research in contextual bandits (\cite{dimakopoulou2017estimation}) shows new policy algorithms that incorporate methods from the causal inference literature in order to reduce bias in the reward estimator and increase its robustness.
In \cite{banditinfra}, LinkedIn, Netflix, Facebook, and Dropbox described engineering systems that can support algorithmic policy making, how they are utilized for their products, and their deepening relationship to attribution, a well known causal inference problem in many industries.
Similarly, fields like artificial intelligence are also dependent on a rich analysis of causal effects.

It is common to find mature software and engineering systems for an experimentation platform
(\cite{fabijan2017benefits}, \cite{kohavi2013online}, \cite{deng2017trustworthy}, \cite{diamantopoulos2019engineering}).
However, it is much less common to find mature software for statistics and causal inference that integrates into such systems.
One of the main challenges is the lack of software dedicated to estimating causal effects that scales well. 
For example, policy algorithms train models over high dimensional feature sets, then use them to evaluate several actions and counterfactuals for different combinations of features, a task that demands a computationally efficient engine. The lack of performant software for such a daunting task creates engineering risk, as well as slow and challenging iteration cycles.

In order to achieve broad adoption of causal inference methods in fields such as experimentation platforms and policy making, the methods need to be general, performant, and robust in software that is easy to use. This requires an interdisciplinary field across causal inference, algorithm design, and numerical computing, which we introduce as \textbf{computational causal inference} (CompCI). The field focuses on software that scales causal inference methods so that they are practical to use broadly in research and in engineering systems.
We aim to improve research agility, as well as software that report treatment effects for online experiments and produce policies that maximize the causal effect on a KPI. By adopting strategies for improving computational performance, causal models can be trained and evaluated efficiently, frequently 30 times faster than off-the-shelf strategies. In addition, there are strategies that can make a single machine scale well, greatly reducing the overhead and maintenance burden in large engineering systems. The combined simplicity and performance reduce friction to apply causal inference in policy making. 

This paper discusses the exact algorithmic and human need for scalable causal effects, specifically in performant, general, and robust estimation. Our introduction of computational causal inference calls for community involvement to solve open challenges in software and methods research for causal inference. We review the state of causal effects software in the industry, with exact references to code. To lead an open discussion, we propose a general framework to structure a software library around causal effects, and specific ways to optimize fitting models and estimating distributions of treatment effects. Early development at Netflix shows promising results. Finally, we share the major challenges we hope the broader community will unite on and solve in the field of computational causal inference.

\section{Engineering Need for Scalable Computation for Causal Effects}

There are two significant classes of engineering systems that motivate the need for performant computation for causal effects: experimentation platforms, and algorithmic policy making engines.

\subsection{Experimentation Platforms}

First, an experimentation platform (XP) that reports on multiple online and controlled experiments needs to be able to estimate causal effects at scale. For each experiment, an XP models a variety of causal effects, for example the common average treatment effect, conditional average treatment effects, and time dynamic treatment effects, seen in Figure 1. These effects help the business understand its user base, different segments in the user base, and how they change over time. The volume of the data demands large amounts of memory, and the estimation of the treatment effects on such volume of data can be overwhelming.

\begin{figure}[h]
\centering
\caption{Types of treatment effects.}
\label{fig:treatment_effects}
\begin{subfigure}{.325\textwidth}
  \centering
  \includegraphics[width=.90\linewidth]{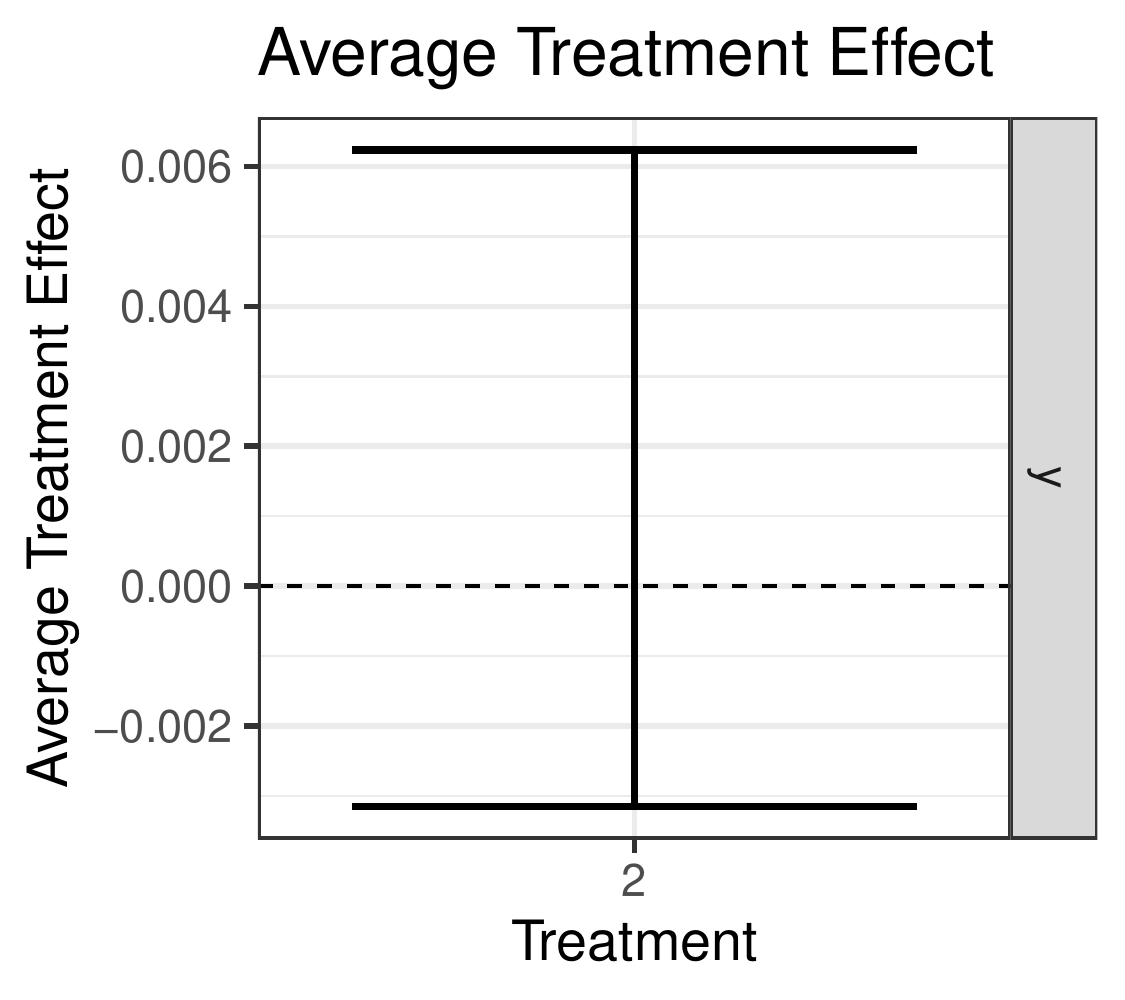}  
  \caption{Average treatment effect.}
\end{subfigure}
\begin{subfigure}{.325\textwidth}
  \centering
  \includegraphics[width=.90\linewidth]{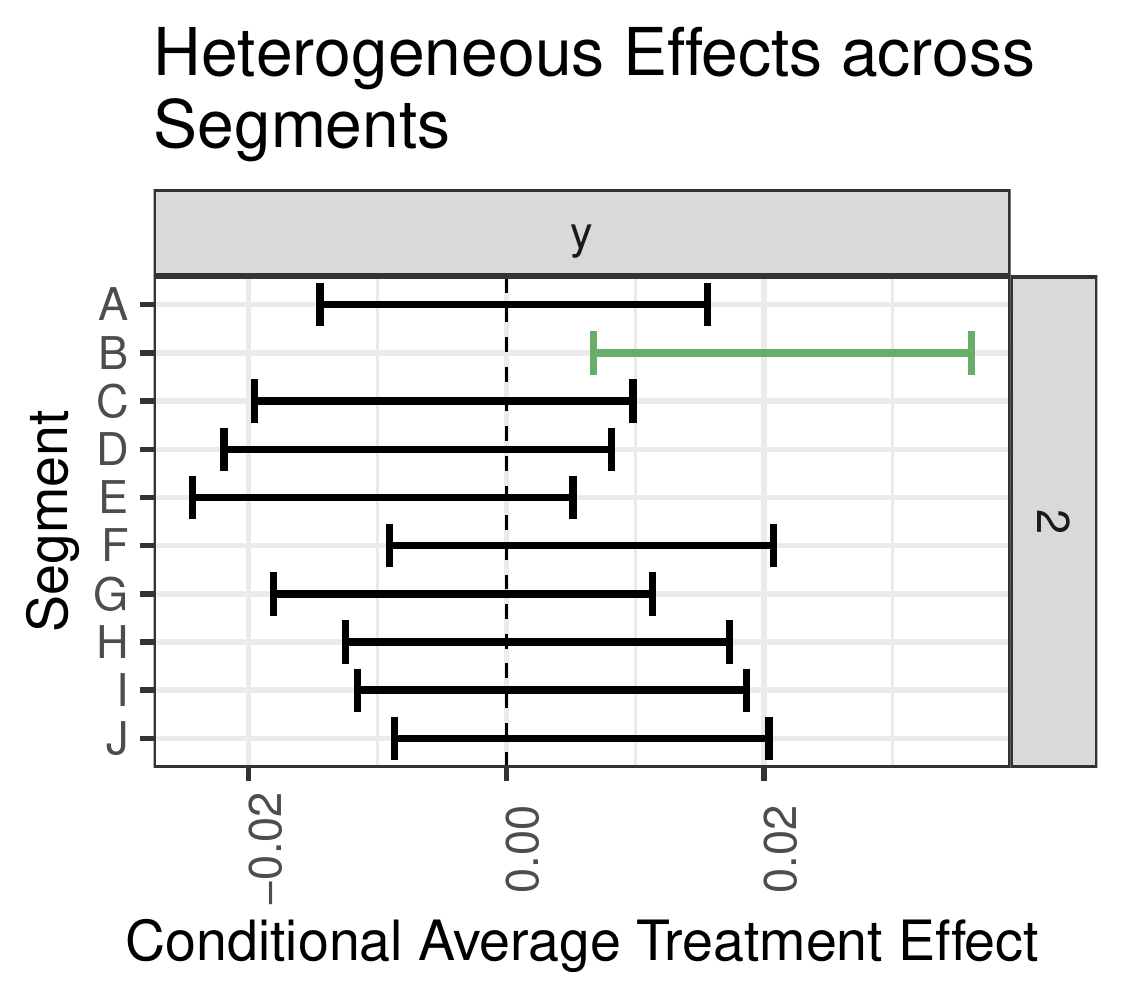}  
\caption{Conditional average treatment effects report the average effect per segment.}
\end{subfigure}
\begin{subfigure}{.325\textwidth}
  \centering
  \includegraphics[width=.90\linewidth]{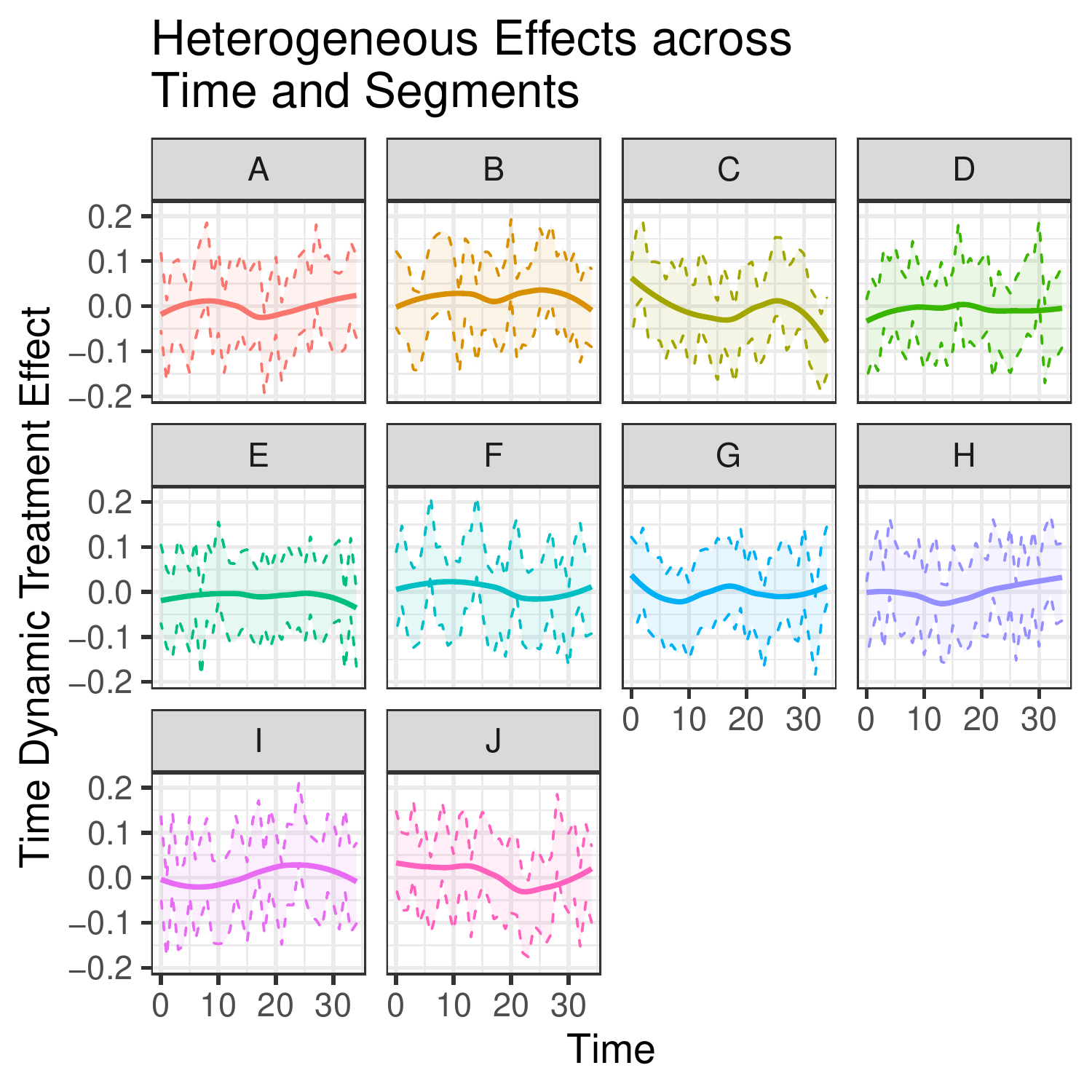}
  \caption{Time dynamic treatment effects report the average effect per segment per unit of time.}
\end{subfigure}
\end{figure}

Ordinary Least Squares (OLS) is usually the foundation for measuring average treatment effects, and extends elegantly into conditional average effects and time dynamic effects (\cite{wong2019efficient}). The first step is to fit OLS, and the second is to query it for counterfactual predictions across all potential outcomes. The computational complexity for fitting OLS is usually $O(np^2)$, where $n$ is the number of observations, and $p$ is the number of features. In practice, an XP can encounter scenarios where $n$ is on the order of $10^8$. The OLS model requires interaction terms in order to estimate conditional average treatment effects, making $p$ large. To measure treatment effects over time, we must observe users for multiple time periods, dramatically increasing the number of observations in the analysis; an analysis of causal effects that took $n = 10^8$ observations can easily become an analysis that takes $n = 3 \cdot 10^9$ or $n = 9 \cdot 10^9$ observations. After fitting such a large OLS model with covariates, $X \in \mathbb{R}^{n \times p}$, and dependent variable, $y$, we must evaluate the model for different potential outcomes. Suppose there are $K$ potential treatments in the set: A = $\{A_1, A_2, \ldots A_K\}$, with $A_1$ being the control experience. The conditional average treatment effects are the conditional differences
\begin{align*}
E[y | A_i, X] - E[y | A_1, X] \qquad \forall i \in \{2 \ldots K\}.
\end{align*}
For each conditional difference, the expectation scores the counterfactual feature matrix of size $n \times p$, where the treatment variable is set to $A_i$. Generating these counterfactual feature matrices and predictions is again a memory and computationally heavy operation. An XP repeats this lengthy exercise for multiple dependent variables and for multiple experiments, culminating in large amounts of computation.

\subsection{Algorithmic Policy Making}

Second, policy algorithms support engineering systems through automated decision making by recommending actions that cause a system to incrementally reach a better state. They have similar computational complexity to that of an XP.
Large applications in product recommendations and artwork have been discussed in \cite{banditinfra},  \cite{artwork_tech_blog} and \cite{artwork}. The setup for policy algorithms begins with $n$ users, and for each user we must decide an action among $K$ actions in $A = \{A_1, A_2, \ldots, A_K\}$. Each user has features, $x$, and each action generates a reward, $R(A, x)$ with respect to a KPI. A deterministic policy, $\pi(x)$, is a function that receives $x$ and returns an action that is believed to generate the optimal reward.
Given the current policy, $\pi_0$, we want to know whether there are alternate policies that achieve a larger reward than $R(\pi_0, x)$, that is we seek to optimize $\pi^*(x) = \argmax_{\pi(x)}  R(\pi(x), x) - R(\pi_0, x)$. This formulation can be thought of as a treatment effect, with $\pi_0$ being the control policy and $\pi(x)$ the treatment policy.

There are many other ways causal effects problems overlap with policy algorithms, such as:

\begin{enumerate}
    \item Identifying the best action that improves over $\pi_0$ requires measuring treatment effects.
    \item Personalized policies seek heterogeneity in treatment effects. Constant treatment effects yield policies that are also constant.
    \item The effect of an action can be a function of time, and can be a function of the actions previously taken. This is similar to analyzing the causal effects of digital ads, which can vary in time and can have compounding or diminishing effects, for example in  \cite{lewis2018incrementality}.
    \item A policy algorithm may suggest to take an action, $\pi^*(x)$. However, the action that is executed may be a different action, or no action at all. This is a case of noncompliance with the treatment, a common phenomenon in many other fields.
    \item A policy algorithm usually assumes that all actions in $A$ can be used with all $n$ users. However, some users may not be qualified for certain actions. Furthermore, the list of actions they are qualified for may change over time. This is similar to modeling causal effects for an at risk group. 
\end{enumerate}

To estimate personalized policies using causal effects, we first fit a causal model. Then, we query it across all possible actions to predict individual treatment effects conditioning on a user's features. One way to build a policy from the individual treatment effects is to find the actions that yield the largest treatment effect. The analysis of causal effects is similar to that in an XP, with the exception that an XP analyzes causal effects retrospectively, and policy algorithms predict causal effects prospectively. Policy algorithms naturally inherit all of the computational complexity that XPs have, and frequently have greater computational complexity than an XP. 

The evaluation of a policy is different than the evaluation of a single treatment in an XP, and introduces greater computational complexity for policy engines. Policy algorithms assign variable treatments to different users, whereas an XP reports effects conditioning on a fixed treatment.
In order to test if the personalized policy, $\pi^*$, is better than $\pi_0$, we can test the hypothesis 

\begin{align*}
 H_0: & \sum_i R(\pi^*(x_i), x_i) - R(\pi_0(x_i), x_i) = 0 \quad \text{against} \\
 H_A: & \sum_i R(\pi^*(x_i), x_i) - R(\pi_0(x_i), x_i) > 0.
\end{align*}
This is similar to offline policy evaluation in \cite{li2011unbiased}, where the policy is evaluated using the sum of the rewards conditioning on $x$. The hypothesis test keeps the reward evaluation general, even in the case when data is autocorrelated. In this case, estimating the distribution of the sum of the treatment effects can be numerically challenging, for example in the case of clustered covariances for OLS (\cite{newey1986simple}). In the more challenging case when a closed form solution for the distribution of the treatment effect does not exist, we may be able to defer to the bootstrap (\cite{efron1994introduction}). However, the bootstrap requires fitting the model multiple times, further highlighting the causal inference and numerical challenges in policy engines.

In summary, experimentation platforms use rich causal effects to inform business strategy. These effects are numerically challenging to estimate. Policy engines are similar to XPs, utilizing causal effects for algorithmic decision making, but have even greater numerical challenges. Computational causal inference will provide a focus on scalable algorithms for causal inference, improving engineering systems for both fields.

\section{Human Productivity Needs Agility}

Computational causal inference's focus on numerical computing delivers agility that enables people to innovate, be productive, and quickly recover from mistakes.
This unique focus in CompCI  simultaneously serves both the engineering needs in the industry as well as the needs to iterate quickly in research.

New industry engineering systems need to be assessed for risk and for their ability to scale before they are deployed. Sustaining an engineering system is a long term commitment, so the system needs to be capable of continuously scaling with the industry for years. Massive computations, such as the ones we outlined for XPs and policy engines, are risky, and it is unclear whether simply acquiring more hardware can solve scaling challenges. The risk causes teams to debate the gains and costs of integrating causal inference into their systems. Often, there are three major costs to a team:

\begin{enumerate}
    \item Instability or latency in the product for the end consumers.
    \item The risk that scaling becomes too expensive in terms of money, hardware, and/or time, and will require a significant redesign in the future. The redesign may include the redesign of other engineering services in the ecosystem.
    \item The associated loss in team productivity due to managing such large computations.
\end{enumerate}
Alternatively, teams may create more scalable, but nuanced, approximations that deviate from rigorous mathematical frameworks in causal inference. Such deviations can create challenges in the future, where it becomes hard to extend, and hard to absorb new innovations from the causal inference field. CompCI preempts the scalability challenges by optimizing the numerical algorithms for causal effects, reducing the risk in developing and maintaining engineering systems that are based on causal inference. Secondly, CompCI's approach allows failed tasks to be restarted and rerun quickly, improving developer operations and maintainability of the system.

Fast computing helps researchers become productive and innovative. First, fast or interactive computing maximizes cognitive flow (\cite{gray2017importance}). Scalable computation that runs on a single machine removes the cognitive overhead of managing distributed environments. Attention becomes concentrated on a single machine that returns results in seconds, facilitating a two way communication between the researcher and the data. This helps the researcher transform thought into code, and results into new thought, ultimately improving productivity. Second, fast computing empowers researchers to innovate by making failure less costly. Innovations are always preceded by a long series of failures, which can bear high mental and emotional costs. The challenges to success can be exacerbated when iterations take hours instead of seconds. Fast software not only saves time for the researcher, it makes it easier to recover from mistakes, lowering psychological barriers to innovate and amplifying the researcher's determination.
To support the possibility of such an experience, we outline computational strategies in section 7 that greatly improve the performance of causal inference software.

Finally, CompCI provides a path for researchers and industry practitioners to use the same software, which is easy to iterate on and can run large problems on a single machine. This is a powerful productivity advantage for XPs and algorithmic policy making, which are receiving significant development from both academic and industry communities.

\section{Community and Involvement}

Computational causal inference is an interdisciplinary field with a broad audience. Its impact benefits engineers in industry who are developing large scale systems, as well as methods researchers. The relationship between the two communities becomes stronger by consolidating software methods, where CompCI software can be deployed in distributed environments, and on a single machine. 
Furthermore, the community is better able to leverage causal inference across wider applications.

There are several unsolved challenges in CompCI, and we believe a community of domain experts is required to solve them.
As an interdisciplinary field, we invite causal inference scientists, experimenters, statisticians, policy makers, algorithms designers, and software engineers to help advance the methods, application, scalability, and deployment of causal effects. As examples, researchers with experience in causal identification with complex data, for example in econometrics, experimenters with experience analyzing imperfect experiments, for example in clinical trials with noncompliers and defiers, and engineers in numerical computing can have a tremendous impact on CompCI software. We discuss examples of major challenges in a later section.

The remainder of this paper describes ways the community can unite to strengthen the CompCI field. We take the initiative to begin the discussion on major topics, with the aim to evolve a solution together with other experts in the community.

\section{State of the Industry}

Many resources have been devoted to improving the performance of machine learned models, and  integrating machine learned models with other engineering systems seamlessly. For instance, Xgboost has been designed from its inception to be a highly performant tree boosting algorithm (\cite{chen2016xgboost}), and contributors have added methods to apply GPU acceleration (\cite{mitchell2017accelerating}). Nvidia has invested multiple iterations of cuDNN (\cite{chetlur2014cudnn}), a library of deep learning primitives. TensorFlow (\cite{abadi2016tensorflow}) has also received much attention on computing (\cite{mo2017performance}, \cite{sergeev2018horovod}, \cite{jia2018improving}). However, fewer resources have been devoted to computational performance and engineering for causal effects.

Spark, Python, and R are large contributors to programmatic access to industry data science. Spark is a distributed computation engine that scales well for data engineering and parts of data science, though there is not much development for causal inference. Python and R have rich ecosystems for modeling causal effects, for example generalized linear models, linear mixed effects models, instrumental variables, matching, propensity scores, doubly robust estimators and regression discontinuity. The R library, grf (\cite{wager2018estimation}), represents cutting edge research that estimates heterogeneous treatment effects using random forests.

At the time of writing, each of these models face scaling challenges. However, combining best practices from numerical computing can greatly enhance scalability. For many problems, the improved performance affords the luxury of developing in single machine computing environments, and makes research and development more agile. Below, we highlight the state of common software implementations for causal effects models.

\subsection{Ordinary Least Squares in StatsModels and R}

\begin{enumerate}
    \item Ordinary Least Squares (OLS) assumes a model in the form $y = \beta_0 + A\beta_1 + X\beta_2 + \varepsilon$ where $\varepsilon \sim N(0, \Sigma)$ and $\Sigma$ is block diagonal. It estimates the parameters using the normal equations, fitting $\hat{\beta} = (M^T M)^{-1} M^T y$ where $M = \begin{bmatrix}
1 & A & X
\end{bmatrix} \in \mathbb{R}^{n \times 1+p+K}$ is the model matrix concatenating a ones vector for the intercept with the treatment variables, $A$, and covariates, $X$. The covariances of the parameters are $cov(\hat{\beta}) = (M^T M)^{-1} M^T \Sigma M (M^T M)^{-1}$.
    \item To fit the model, both StatsModels and R will convert a dataframe to a model matrix, $M$, then run a matrix decomposition on $M$ (\cite{ols_sm} and \cite{ols_r}). For example, using the SVD or QR decompositions.
    \item StatsModels fits this model using numpy arrays, which are used to represent a dense matrix. R uses numeric vectors. Both data structures are dense, so neither language is optimized for storage when the feature matrix has many zeroes, nor are they optimized for sparse linear algebra.
    \item We can compute the difference in counterfactuals, $E[y | A = A_i, X] - E[y | A = A_1, X]$, by naively constructing two counterfactual matrices, $M(A = A_i)$ and $M(A = A_1)$. In these two matrices we set the treatment variable to $A_i$ and $A_1$ respectively. Then the treatment effect is the difference in the predicted counterfactual outcome, $M(A = A_i)\hat{\beta} - M(A = A_1)\hat{\beta}$. If there are $K$ actions to evaluate, there would be $K$ counterfactual matrices to generate, all of which would be dense matrices and would suffer from inefficient storage and linear algebra.
\end{enumerate}
    
With these constraints it is difficult to use linear models to iterate and find treatment effect heterogeneity. On a problem with $n=10^7$ observations, $p=200$ features, a dense linear algebra solver spends 30 minutes to compute 1000 CATEs.

\subsection{2 Stage Least Squares in StatsModels and R's AER}

Two stage least squares (\cite{angrist1995two}) is a model that can estimate a local average treatment effect when the treatment variable is endogeneous. The model estimates the system

\begin{align}
    y &= \beta_0 + A\beta_1 + X\beta_2 + \varepsilon, \\
    A &= \gamma_0 + Z \gamma_1 + X\gamma_2 + \nu.
\end{align}
This is estimated by first fitting an OLS model to the first stage: $A = \gamma_0 + Z \gamma_1 + X\gamma_2 + \nu$. The fitted values, $\hat{A} = \hat{\gamma}_0 + Z\hat{\gamma}_1 + X\hat{\gamma}_2$, are used to estimate the second stage model: $y = \beta_0 + \hat{A}\beta_1 + X\beta_2 + \varepsilon$. This common implementation is also not scalable for sparse data. When $A$ and $Z$ are sparse, the first stage can be solved efficiently with sparse linear algebra, but the implementations in StatsModels and AER (\cite{2sls_python} and \cite{2sls_aer}) use dense algebra. Furthermore, the design of the algorithm relies on materializing $\hat{A}$ in memory, which is dense because $\hat{\gamma}$ is dense. By fitting two naive OLS models this implementation forces the second stage to be fit using dense methods, even when $A$ and $Z$ are originally sparse.

\subsection{Generalized Random Forests in R}

Generalized Random Forests (grf) (\cite{athey2019generalized}) is a rich model based on random forests that can estimate heterogeneity in treatment effects. The software has a highly optimized C++ core and was designed with scalability in mind. However, heterogeneous treatment effect estimation is inherently complex, making it difficult for large problems. In practice, evaluating causal effects with $K$ treatments and $m$ KPIs requires $K \cdot m$ calls to grf (\cite{multicausalforest}). A single call to grf using a problem with $n = 2 \cdot 10^6$, $p = 10$ and $\textit{num.trees} = 4000$ takes 2 hours to fit the ensemble with 32 active cores.

\section{Developing a Software Framework for Measuring Treatment Effects}

In order to create leverage, computational causal inference needs to generalize a software framework for computing different causal effects from different models, then needs to optimize that framework.

We begin our discussion by using the potential outcomes framework to measure treatment effects. First, we assume a model for the relationship between a KPI, $y$, a treatment variable, $A$, and other exogeneous features, $X$. For simplicity, let $A$ be a binary indicator variable where $A = 1$ represents a user receiving the treatment experience, and $A = 0$ the control experience. We can estimate the difference between providing the treatment experience to the user, and not providing the treatment experience, by taking the difference in conditional expectations $$E[y | A = 1, X] - E[y | A = 0, X].$$
If a model, $y = f(A, X)$, has already been assumed apriori, then this definition for the conditional treatment effect is simple.

However, experimenting across many models is difficult. There are many models that can estimate $E[y | A, X]$. Each of these has different requirements for the input data, each has different options, each has a different estimation strategy, and each has a different integration strategy into engineering systems.
CompCI needs to generalize a software framework beyond the potential outcomes framework.

Design patterns in machine learning frameworks are leading examples for how software can democratize and provide programmatic access across many models. First, these frameworks have several built-in models, but also expose an API to define an arbitrary model as the minimization of a cost function. The frameworks then apply generic numerical optimization functions to these arbitrary models. The TensorFlow tutorial page (\cite{tensorflowtutorial}) shows simple and composable code that allows the user to specify the form of a model, then train it without needing to derive an estimation strategy.

\begin{lstlisting}[language=Python]

model = keras.Sequential([
    keras.layers.Flatten(input_shape=(28, 28)),
    keras.layers.Dense(128, activation="relu"),
    keras.layers.Dense(10)
])
model.compile(
    optimizer="adam",
    loss=tf.keras.losses.SparseCategoricalCrossentropy(from_logits=True),
    metrics=["accuracy"]
)
model.fit(train_images, train_labels, epochs=10)

\end{lstlisting}
The framework then provides a single entrypoint to make predictions using an arbitrary model.

\begin{lstlisting}[language=Python]

probability_model = tf.keras.Sequential([
    model, 
    tf.keras.layers.Softmax()
])
predictions = probability_model.predict(test_images)

\end{lstlisting}

This simplicity makes integration with other engineering systems seamless; deploying a change to the form of the model automatically changes the estimation strategy and predictions on new features, and is a single point of change for a developer.

CompCI seeks similar software that can generalize the software framing of causal inference problems, create a structured framework for computing causal effects, and make software deployment simple. Frameworks like TensorFlow already simplify the process of training an arbitrary model. After a model is trained, the conditional treatment effect is the difference in conditional expectations comparing the treatment experience to the control experience. However, causal inference problems have two additional layers of complexity that demand a powerful software framework. 

First, there are conditions on when the conditional difference can be safely interpreted as a causal effect. Say the model is parameterized by $\theta_f$ so that $y = f(A, X; \theta_f)$. If the conditional difference is written as a function $g(A, X; \theta_g)$, with $\theta_g \subseteq \theta_f$, it is a causal effect if $\theta_g$ is identified. In parametric models, a parameter $\theta^*$ is identified (\cite{koopmans1949identification}) if it has an estimator $\hat{\theta}^*$ that is consistent for $\theta^*$, and the convergence of $\hat{\theta}^*$ to $\theta^*$ does not depend on other parameters. Identification is a property that varies by model, making it challenging for a framework to detect whether an arbitrary model has parameters that can be identified.
In most cases it requires declaring assumptions about the data generating process, which should be made understandable to the framework in order to provide safe estimation of treatment effects.
After determining a collection of models that have causal identification, the second software challenge is to estimate the distribution of the treatment effect from an arbitrary model. A possible solution to this is to estimate the distribution through the bootstrap (\cite{efron1994introduction}). Together, arbitrary model training, safe identification of causal effects, the bootstrap, and the potential outcomes framework can create a general framework for computing treatment effects that can be leveraged in an experimentation platform and an algorithmic policy making engine.

\section{Optimizing Computational Performance}

In addition to providing a general framework for measuring causal effects, CompCI software must be scalable and performant. Without this second quality, the software will still be difficult to integrate into engineering systems. The previous State of the Industry section shows common demands in computing: 
\begin{enumerate}
    \item Optimize for sparse data.
    \item Efficiently build, and predict with, counterfactual feature matrices.
    \item Vectorize or parallelize for multiple KPIs and multiple treatments.
    \item Estimate the distribution of the treatment effect efficiently.
\end{enumerate}
Below we provide an overview of strategies that can address these themes; when combined these strategies yield significant performance improvements to causal inference software. Greater details on each strategy can be found in our related CompCI papers.

\subsection{Compress Data Volume}

No matter the form of the causal model, a large volume of observations is a significant contributor to computational complexity. For a certain class of models, it is possible to reduce the volume of data while still preserving estimators for average effects, conditional average effects, and time dynamic effects. Since these effects are averages of counterfactual predictions, it is possible to aggregate data at the beginning of the modeling process and still return estimates of the treatment effects.

In the simplest example, the two-sample t-test can be recast to the simple least squares model $y = \alpha + A\beta_1 + \varepsilon$. This models $y$'s conditional means per unique condition: for the treatment group, $A = 1$, and the control group, $A = 0$. The $\beta_1$ coefficient represents the average difference between the groups. The model can be estimated using size $n$ arrays, or its data can be aggregated to just the mean and variances of the treatment and control group. This simple example leads to an elegant generalization for data compression using conditional sufficient statistics. A larger linear model that conditions on more features, $X$, can operate on matrices with $n$ rows, or its data can be aggregated to just the mean and variance per unique condition. A modification to the standard OLS algorithm allows linear models to be estimated on aggregated data without loss. Strategies for compressing for average effects, conditional average effects, and time dynamic effects are discussed in a related CompCI paper (\cite{yoco}).

\subsection{Optimize the Software Stack for Sparse Data}

There are three opportunities for sparse data optimization in the modeling stack. First, the creation of a feature matrix using data can be optimized for sparse features. Second, model training can be optimized for sparse algebra. Third, estimating the treatment effects can also be optimized for sparse algebra.

Data is frequently retrieved from a data warehouse in the form of a dataframe, which is then encoded as a feature matrix. A significant optimization is to convert data from warehouses directly into feature matrices, without constructing dataframes. For example, parquet files are serialized with the arrow format (\cite{arrow}) for columnar storage, and are aligned with the common columnar storage formats for matrices. Software that constructs a feature matrix from a parquet file, eliminating overhead from dataframes, would have great impact.

Feature matrices frequently contain many zeroes, especially when $A$ or $X$ contain categorical variables that are one-hot encoded. The software library creating the feature matrix, $M$, must be optimized so that the feature matrix is sparse, which will decrease storage costs and will improve subsequent modeling steps. This is done in SparkML (\cite{sparkml_features}), and R's Matrix library (\cite{matrix_r}). After constructing $M$, estimating the model with $M$ should optimize any linear algebra operations as sparse linear algebra. For example, high performance computing for linear models can be achieved using sparse feature matrices and sparse cholesky decompositions in Eigen (\cite{eigenweb}).

Finally, the estimation of treatment effects through the difference in conditional expectations should also optimize for the creation of sparse feature matrices and sparse linear algebra.
Because the difference in conditional expectations holds all $X$ variables constant, and only vary the $A$ variable, the conditional difference can be represented as operations on a sparse matrix.

\subsection{Leverage Structure in Data and Models}

Estimating conditional average treatment effects requires constructing counterfactual feature matrices to simulate the difference between treatment and control experiences. Even with sparse data optimizations this can be a large computational task. In a related CompCI paper, \cite{wong2019efficient}, shows it is possible to leverage structure in linear models to estimate conditional average effects across multiple treatments, without allocating memory for large counterfactual matrices, by reusing the model matrix used for training.

\subsection{Vectorize for Multiple Treatments and Multiple KPIs}

In both applications for an experimentation platform and an algorithmic policy making engine, there are multiple treatments to consider. The experimenter wants to analyze the effect of each treatment and identify the best one. There may also be multiple KPIs that are used to determine the best treatment. The evaluation of many different treatment effects for many outcomes can usually be done in a vectorized way, where computational overhead is minimized and iteration over KPIs and causal effects have minimal incremental computation. For example, OLS estimates a set of parameters that analyze multiple treatments simultaneously by one-hot encoding the treatment variable and using heteroskedasticity-consistent covariances (\cite{eicker1967heteroskedasticity,huber1967heteroskedasticity,white1980heteroskedasticity}). The normal equations for OLS can be extended to analyze multiple KPIs simultaneously with minimal incremental computation by computing $\hat{\beta} = (M^T M)^{-1} M^T Y$, where $Y$ is a matrix of KPIs that share a common $(M^T M)^{-1}$.

\subsection{Bag of Little Bootstraps}

We can estimate the sampling distribution of treatment effects generically using the bootstrap. To do this at scale, we may implement the bag of little bootstraps (\cite{kleiner2014scalable}), an efficient way to compute the bootstrap by dividing the data into multiple small partitions, then bootstrapping within each partition. This method can be run in parallel and is scalable. Furthermore, it is generic and can be applied to models without knowing specific properties apriori.

By integrating into a general framework for measuring treatment effects, the bag of little bootstraps becomes an engineering abstraction that allows developers to focus on causal identification and parameter estimation, without having to write specialized functions for estimating the distribution of treatment effects. It is a fundamental component to create a simple and unified framework.

\subsection{Strategies from High Performance Numerical Computing}

In addition to the above strategies, CompCI should leverage conventional wisdom from high performance numerical computing.


\subsubsection{Minimize Memory Allocations}

Memory allocations and deallocations can consume a significant amount of time in numerical computing. For example, one software implementation for computing the variance on a vector can use the sum of the vector, and the sum of its squared elements. Allocating a vector to represent its squared elements would be inefficient because the vector will be reduced through the sum function. Numerical algorithms should design for the end target in mind, and minimize memory allocations along the way.

Conditional average treatment effects can be thought of as the average treatment effect among a subset of users. This can be computed by taking the subset of the feature matrix, computing counterfactual predictions, then taking the difference. To minimize memory allocations, the subset of the feature matrix should not create a deep copy of the data, it should be a view of the original data. Among linear models, another implementation is to never subset the feature matrix directly, instead multiply it by a vector of ones and zeros to select the observations that belong to a particular subset.

\subsubsection{Optimize Memory Access}

Computations can be optimized by making use of cache memory. One way to do this is to load data so that it is accessed sequentially, improving spatial locality. For example, when computing the treatment effect for a specific subpopulation where $X = x^*$, spatial locality can be improved by loading data that is sorted apriori so that users with $X = x^*$ are in consecutive memory blocks.

\section{Potential for the Future}

Netflix's experimentation platform has been investing heavily in CompCI software. Estimating treatment effects on large data with their software is approximately 30 times more performant than off-the-shelf libraries. Estimating 1000 conditional average effects with $n = 5 \cdot 10^7$ and $p \approx 200$ returns in 10 seconds on a single machine. In an extremely large problem with $n = 5 \cdot 10^8$ and $p \approx 10000$, conditional time dynamic treatment effects on a longitudinal dataset with clustered covariances were computed in one hour on a single machine. This has made integrating research on heterogeneous treatment effects and time dynamic treatment effects into the XP more scalable and less risky. Furthermore, single-machine software for modeling has greatly simplified engineering management in the XP, and has also created a simpler path for elevating new research.

\section{Open Challenges}

As an emerging field, computational causal inference has a plethora of challenges, including generalizability of causal inference, numerical computation, statistics and probability theory, software design, and applications of causal inference.
We believe experts from different domains are needed to solve remaining problems in CompCI.
Below is a list of significant challenges in this new field that we wish to open to the community.

\begin{enumerate}
    \item Structuring software to enable detection of causal identification can create safe, scalable, and programmatic access to causal effects. It is a challenge because the framework must know some properties of the models implemented. Previous literature exists for identification in causal graphs where the data generating process can be represented as a directed and acyclic graph (DAG) (\cite{pearl1995causal} and \cite{tian2002general}). In such a graph, data are represented as nodes, and causal effects are represented as edges. The backdoor criterion allows an experimenter to query a DAG and know if the causal effect on $y$ due to $X$ is identified. This can shift the responsibility of causal identification from the model to the data, making it easy for a framework for causal effects to adopt. However, it is unclear how identification in a DAG extends to causal graphs when relationships can be cyclic, which is a common phenomenon in machine learned systems that have immediate feedback loops.
    
    \item One of the ways machine learning has been democratized is the fact that most cost functions for models are differentiable, and can be optimized through stochastic gradient descent (SGD) (\cite{bottou2010large}). This leads to minimal friction for developers that want to create, or iterate on, a model: given only a functional form and a cost function, a model can be estimated and predictions can be generated. Similarly, we could structure CompCI software around the class of models with causal identification that are also differentiable, then train them generically using SGD. However, SGD requires hyperparameter tuning, and it is unclear how the risk of poor convergence affect bias in the causal effects estimators. In particular, we do not have a robust understanding of the convergence specifically on the causal parameters of interest.
    
    \item Tracing causal effects for $n$ users over $T$ time periods requires a dataset to grow to $n \cdot T$ observations. When computing the distribution of the treatment effect, a model must also acknowledge the autocorrelation in the data. This becomes both statistically and computationally challenging. A solution to this will prevent overconfidence in the estimates, and is also relevant for offline policy evaluation, since most methods make a simplifying assumption that the data is not autocorrelated.
    
    \item In algorithmic policy making, policy makers may hypothesize that additional treatments have compounding or diminishing returns. While the first treatment can be randomized, experiencing a second treatment is not necessarily random; for example, there can be a selection bias in who returns for a second round of treatment. Developing software that can determine identification for marginal treatment effects is hard, but will create safer software, and can be used in common situations where a user returns for multiple treatments. Instrumental variable methods can sometimes be used to estimate marginal effects when the first round of treatment is randomized but subsequent rounds are not, however these methods are difficult to scale. A solution for this is a significant contribution to attribution analysis.
    
    \item The set of treatments that a user can experience can change over time and with context. When learning from historic data, we must distinguish between a user not experiencing treatment $A$, from treatment $A$ not being available at that time. If we have the ability to randomize the data, we must also understand how the probability of experiencing treatment $A$ varied with context, availability, and time. Software that understands conditional randomization, and that the conditions change in time, are extremely important for policy engines that evolve with a business over time.

\end{enumerate}

\section{Conclusion}

Computational causal inference, CompCI, is an interdisciplinary field across causal inference, algorithms design, and numerical computing. CompCI aims to unite the community of engineers, scientists, experimenters, statisticians and many others to develop mature software for causal inference. The field addresses engineering needs and human needs for scalability, and directly benefits the deepening relationship between experimentation and personalization in products and algorithms. For example, experimentation platforms and policy engines both use causal inference to drive innovation, automation, and personalization in a company.

We proposed the design for CompCI software follow design principles from machine learning and the potential outcomes framework, which would separate the framing and identification of causal effects from its estimation strategy. By doing so we can develop a general and public framework for a wide variety of causal effects, then optimize its internal numerical engine privately. This software design would allow users to estimate causal effects by simply specifying the form of a model, and assumptions of the data that can be used to determine if effects are identified. Furthermore, we shared details of numerical computing that are crucial for improving the performance of causal inference algorithms. Mature CompCI software will minimize the amount of friction in researching, developing and applying causal effects to large datasets.

We have started the discussion and made a large investment in increasing the performance of causal inference algorithms.
Other fields in physical sciences, social sciences, experimentation, statistics and engineering have local expertise that can be contributed to this interdisciplinary field.
As an emerging field, we left several open challenges as a call for others to contribute and help develop this community.

\printbibliography

\end{document}